\documentclass[pra,twocolumn,superscriptaddress,a4paper,nofootinbib]{revtex4}
\usepackage{amsmath}
\usepackage{amsfonts}

\begin{document}
\title{Relativistically invariant quantum information}
\author{Stephen D. Bartlett}
\email{bartlett@physics.uq.edu.au}
\affiliation{School of Physical Sciences, The University of Queensland,
  Queensland 4072, Australia}
\author{Daniel R. Terno}
\email{dterno@perimeterinstitute.ca}
\affiliation{Perimeter Institute for Theoretical Physics, Waterloo,
  Ontario N2J 2W9, Canada}
\date{4 January 2005}

\begin{abstract}
  We show that quantum information can be encoded into entangled
  states of multiple indistinguishable particles in such a way that
  any inertial observer can prepare, manipulate, or measure the
  encoded state independent of their Lorentz reference frame.  Such
  relativistically invariant quantum information is free of the
  difficulties associated with encoding into spin or other degrees of
  freedom in a relativistic context.
\end{abstract}

\maketitle

\section{Introduction}

Information encoded into the states of quantum systems allows for
powerful new computational and communication tasks~\cite{Nie00}.  It
is perhaps in situations involving extremely long distances that
quantum information will find its most useful applications: quantum
teleportation~\cite{Ben93}, entanglement-enhanced
communication~\cite{Gio03}, quantum clock
synchonization~\cite{Joz00,Gio01} and reference frame
alignment~\cite{Per01,Bag01,Lin03,Chi04,Bag04}, and quantum-enhanced
global positioning~\cite{Gio01} are just some of the ways that quantum
physics offers an advantage over classical methods.  In these
long-distance situations, relativistic effects can be expected to
arise~\cite{Per03}.  Consider the canonical example of a qubit encoded
into the angular momentum state of a massive spin-1/2 particle. The
spin entropy, which quantifies the purity of the encoded information,
is not a covariant quantity~\cite{pt:03}: under a Lorentz
transformation, the spin state becomes entangled with the momentum of
the particle. The effect of Lorentz transformations is to
\emph{decohere} the qubit, reducing the applicability of such systems
to perform quantum information processing tasks in a relativistic
setting~\cite{Per03,pt:03}.  Photon polarization qubits behave
similarly, with additional effects arising from the transversality of
polarization~\cite{Per03,pt:03}.

We show that \emph{relativistically invariant} quantum information can
be encoded into entangled states of multiple, indistinguishable
particles.  This encoding allows any inertial observer to prepare and
manipulate quantum information in a way that is independent of their
particular frame of reference.  In particular, two observers can share
entanglement and thus perform any quantum information processing task
(teleportation, communication, etc.)\ without sharing a reference
frame.  We develop such encodings by showing that, under a general
Lorentz transformation $\Lambda_{AB}$, the spin state of a particle
will be transformed due to three distinct effects: (i) a Wigner
rotation due to the Lorentz boost $\Lambda_{BA}$, which occurs even
for momentum eigenstates, (ii) a decoherence due to the entangling of
the spin and momentum under the Lorentz transformation $\Lambda_{AB}$
because the particle is not in a momentum eigenstate, and (iii) a
decoherence due to Bob's lack of knowledge about the transformation
relating his reference frame to Alice's frame.  To construct encodings
that are protected from decoherence of the forms (i) and (iii), we
construct states of multiple indistingishable particles with
well-defined momenta and use the techniques of noiseless
subsystems~\cite{Zan97,Kni00}.  We begin by considering massive
spin-1/2 particles; massless photons are then given a separate
treatment.

\section{Single spin-$1/2$ particle}

Consider two inertial observers, Alice and Bob, who wish to exchange
spin-1/2 particles (e.g., protons) for the purposes of some quantum
information processing task.  First, we consider the exchange of a
single particle and outline the associated difficulties.  To fix our
notation, momentum eigenstates $|\mathbf{0}m\rangle$ with
$m=\pm\frac{1}{2}$ of a single spin-1/2 particle in the rest frame
($\mathbf{p}=\mathbf{0}$),
%are defined by~\cite{wei},
%\begin{gather}
%  \label{eq:MomEigenstate}
%  P^\mu |\mathbf{0}m\rangle = p_0^\mu|\mathbf{0}m\rangle \, , \\
% \mathbf{J}^2 |\mathbf{0}m\rangle = \tfrac{3}{4}|\mathbf{0}m\rangle
%  \, , \quad
%  J_z |\mathbf{0}m\rangle = m|\mathbf{0}m\rangle \, ,
%\end{gather}
%and
are given in a boosted frame as $|\mathbf{p}m\rangle =
L(\mathbf{\xi}_{\mathbf{p}})|\mathbf{0}m \rangle$ for
$L(\mathbf{\xi}_{\mathbf{p}})$ a pure Lorentz boost.  The Lorentz
transformation $\Lambda$ acts via the one-particle representation
$T_1$ as
\begin{equation}
  \label{eq:SingleParticleLorentz}
  T_1(\Lambda)|\mathbf{p}m\rangle =
  \sum_{m'}|(\Lambda\mathbf{p})m'\rangle
  D^{1/2}_{m',m}(\Omega(\Lambda,\mathbf{p})) \, ,
\end{equation}
where $\Omega(\Lambda,\mathbf{p}) =
L(\mathbf{\xi}_{\Lambda\mathbf{p}})^{-1} T_1(\Lambda)
L(\mathbf{\xi}_{\mathbf{p}}) \in$ SO(3) is a Wigner rotation, and
$D^{1/2}_{m',m}(\Omega)$ is its the spin-1/2 representation. Thus, on
the spin degrees of freedom, the Lorentz transformation acts as a
rotation.

Let Alice prepare a single spin-1/2 particle in a state $\rho$
with respect to her reference frame.  This state cannot be an
(unphysical) eigenstate of momentum~\cite{Per03}; the spatial
state of the particle could be prepared, for example, in a
coherent state of minimum uncertainty in both position and
momentum. A generic pure state for a single particle is given in
terms of the basis above by
\begin{equation}
  \label{wave}
  |\Psi\rangle_1=\sum_m \int_{-\infty}^\infty
  \psi_m(\mathbf{p})|\mathbf{p}m\rangle \,{\rm d}\mu(\mathbf{p})\, ,
\end{equation}
where ${\rm d}\mu(\mathbf{p})=(2\pi)^{-3}(2p^0)^{-1}{\rm
  d}^3\mathbf{p}$.  To encode a qubit into this particle, Alice may
prepare the spin of this particle in an arbitrary encoded state
uncoupled (in a product state) with a localized spatial state,
i.e.,
\begin{equation}
  \label{rest}
  |\Psi\rangle_1= \begin{pmatrix} \zeta \\ \eta \end{pmatrix} \int
   \psi(\mathbf{p})|\mathbf{p}\rangle\, {\rm d}\mu(\mathbf{p})\,,
\end{equation}
where we take the wave function $\psi$ to be concentrated near zero
momentum and with a characteristic spread $\Delta$; i.e., to be of the
Gaussian form
$\psi(\mathbf{p})=C\exp\left(-\mathbf{p}^2/2\Delta^2\right)$, where
$C$ is a normalization constant.  The reduced density matrix for the
spin component of this state in Alice's frame is
\begin{equation}
  \label{eq:SpinReducedRho}
  \rho_1=\begin{pmatrix} |\zeta|^2 & \zeta\eta^* \\ \zeta^*\eta &
  |\eta|^2 \end{pmatrix}\,,
\end{equation}
and in this frame is independent of the form of
$\psi(\mathbf{p})$.

Now consider the state of this particle as described by another
inertial observer, Bob.  Let $\Lambda_{BA}$ be the element of the
Lorentz group that relates Bob's inertial frame $B$ to Alice's frame
$A$; Bob thus assigns the transformed state
$T_1(\Lambda_{BA})|\Psi\rangle_1$ to the particle.  Even if Bob has
the perfect knowledge of the relative orientation and velocity of his
reference frame with respect to Alice's, the reduced density matrix
for the spin degrees of freedom of this qubit decoheres~\cite{Per03}.
For example, if the Lorentz transformation $\Lambda_{BA}$ is a pure
boost along the $z$-axis to the velocity $v$, the effective state
transformation is \cite{pt:03}
\begin{equation}
  \rho'_1(v) \approx (1-\tfrac{1}{4}\Gamma^2)\rho_1 + \tfrac{1}{8}\Gamma^2
   (\sigma_x\rho_1\sigma_x+\sigma_y\rho_1\sigma_y)\,,
  \label{chanex}
\end{equation}
where $\Gamma=(1-\sqrt{1-v^2})\Delta/v$.  As this decoherence is an
artifact of Bob's different frame, it is in principle possible for Bob
to overcome this decoherence if he has perfect knowledge of the
relation (i.e., the Lorentz transformation $\Lambda_{AB}$) that
relates his frame to Alice's by altering his frame or performing an
appropriate transformation on the state.

However, if Bob does not \emph{know} this relation, the decohering
effects are much more significant.  He represents the state of the
system as a mixture over all possible Lorentz transformations that
could relate Alice's frame to his.  Specifically, we would represent
the state of the particle as
\begin{equation}
  \label{eq:MixOverAllLorentz}
  \mathcal{E}_1(|\Psi\rangle_1\langle\Psi|) = \int {\rm d}\Lambda\,
  f(\Lambda)T_1(\Lambda)|\Psi\rangle_1\langle\Psi| T_1(\Lambda)^\dag \, ,
\end{equation}
where the integration is over the entire Lorentz group, ${\rm
  d}\Lambda$ is its Haar measure and $f(\Lambda)$ describes Bob's
prior estimate of the Lorentz transformation relating the
systems~\footnote{Because the Lorentz group is non-compact, one must
  take care with using the group-invariant measure c.f.~\cite{wkt}.
  The probability distribution $f(\Lambda)$ not only represents Bob's
  knowledge, but also makes the integral converge.}. Viewing the
quantum state $|\Psi\rangle_1$ as a ``catalogue'' of predictions for
the outcomes of future measurements on the particle, the process
$\mathcal{E}_1$ describes the loss of predictive power by Bob due to
his lack of knowledge about the reference frame in which the state of
the particle was prepared~\cite{Bar03}.  It is useful to view the
superoperator $\mathcal{E}_1$ as a form of decoherence.  Rather than
describing an interaction with an environment, this decoherence
represents the resulting decrease in Bob's predictive capacity due to
his lack of knowledge.

Consider the action of this decoherence on the reduced density matrix
$\rho_1$ of Eq.~(\ref{eq:SpinReducedRho}) for the spin component of
a single spin-1/2 particle.  While the Lorentz group acts via
Eq.~(\ref{eq:SingleParticleLorentz}) on each momentum component as the
spin-1/2 representation $D^{1/2}$ of the rotation group, an effective
transformation for the reduced density matrix of the state
(\ref{rest}) involves averaging over different noisy quantum channels.
For example, if the transformation relating Alice's frame to Bob's is
known to be a pure boost along the $z$-axis but the amount of boost
(i.e., $v$) is unknown and described by a distribution $p(v)$, then
the effective transformation on the reduced density matrix for the
spin component is
\begin{equation}
  \label{eq:MixOverBoosts}
  \mathcal{E}_1^{\rm boost}(\rho_1) = \int {\rm d}v\, p(v)
  \rho'_1(v)\, ,
\end{equation}
where $\rho'_1(v)$ is given in Eq.~(\ref{chanex}).  On the other hand,
the lack of knowledge of the relative orientation of the reference
frames alone is sufficient to completely decohere Bob's
qubit~\cite{Bar03}.  Thus, the decoherence due to entanglement between
spin and momentum and the lack of knowledge about the relative motion
cannot make matters worse, and the total decoherence on the reduced
density matrix for the spin component of a single particle is
\begin{equation}
  \label{eq:MixOverAllRot}
  \mathcal{E}_1(\rho_1) = \int {\rm d}\Omega\,
  D^{1/2}(\Omega)\rho_1 D^{1/2}(\Omega)^\dag =\tfrac{1}{2}I \, ,
\end{equation}
where $\Omega \in\, $SO(3) is a rotation, integration is over the
entire group SO(3), and $\tfrac{1}{2}I$ is the completely mixed
density operator on the spin subsystem.  The spin state of the
particle is decohered in Bob's frame to the completely mixed state,
and thus no quantum information can be conveyed to Bob by encoding
into the spin of a single particle.  
%When the relative orientation of
%frames is known, but the relative velocity is not and/or the effects
%of spin-momentum entanglement are taken into account, Bob's density
%matrix depends both on $\psi(\mathbf{p})$ and $f(\Lambda)$. 
This result proves that Alice and Bob cannot share spin entanglement
through the exchange of a single spin-1/2 particle without first
sharing a reference frame.

%The dependence of the resulting density matrix on the momentum
%wavefunction is a signature of Lorentz boosts.  Hence unlike the
%situation when decoherence is due to unknown relative orientation of
%the reference frames, the methods of Sec.~IV give only an approximate
%solution to the decoherence problem.  We return to this point in
%Sec.~VI.
% We note
%that Bob may perform a measurement on the particle in an attempt
%to gain information about the frame in which it was prepared;
%however, such a measurement necessarily disturbs the state in an
%unpredictable way.

\section{Creating distinguishable qubits from indistinguishable
  particles}

As we will show, it is possible to use entangled states
of multiple particles to combat the deleterious effects of this
decoherence. However, first we must demonstrate that it is possible to
use elementary \emph{indistinguishable} particles as
\emph{distinguishable} qubits through an appropriate preparation of
their spatial wavefunctions.  Consider the states of $N$ identical
particles.  To use these particles as qubits to encode quantum
information, they must be prepared in such a way that they are (i)
distinguishable and (ii) relatively localized and at rest with respect
to each other, so that joint (entangling) operations such as
preparations and measurements can be performed on them.  These
conditions are mutually exclusive at first glance, but by
%: for the particles
%to all be at rest with respect to each other, they must all be in the
%eigenstate of zero momentum with respect to some frame, and thus are
%indistinguishable because they are all in the same spatial state.  By
preparing particles in minimum-uncertainty states that are
well-localized (making them distinguishable) and with a sharp common
momentum, we will show that these conditions can be satisfied.

Consider a translation of a single particle state $|\Psi\rangle_1$
of Eq.~(\ref{rest}), $|\Psi_a\rangle_1=e^{-{\rm
i}aP_z}|\Psi\rangle_1$,
%\begin{equation}
%  |\Psi_a\rangle_1=e^{-{\rm i}aP_z}|\Psi\rangle_1=
%  \begin{pmatrix} \zeta \\ \eta \end{pmatrix} \int e^{-{\rm i}p_za}
%  \psi(\mathbf{p})|\mathbf{p}\rangle\, {\rm d}\mu(\mathbf{p})\, ,
%\end{equation}
where  we arbitrarily choose the translation to be  along the
$z$-axis.  The overlap between two one-particle states serves as a
guide to their distinguishability;
\begin{equation}
  {}_1\langle\Psi|\Psi_a\rangle_1=C^2\int
  {\rm d}\mu(\mathbf{p})\, {\rm e}^{-\mathbf{p}^2/\Delta^2} {\rm
  e}^{-{\rm i}p_za/\hbar} \,,
\end{equation}
which should be small.  Because $\Delta\ll mc$, we expand the energy
as $E=mc^2(1+p^2/2mc^2+\ldots)$ and obtain $
{}_1\langle\Psi|\Psi_a\rangle_1 \propto \exp(-a^2\Delta^2/4\hbar^2)$.
Thus, the condition for distinguishability is
$a\gg\lambdabar/\epsilon$, where $\Delta\equiv\epsilon mc$ and
$\lambdabar=mc/\hbar$ is Compton wavelength of the particle.  Now we
apply our second condition: that the particles should be nearly at
rest in Alice's frame, i.e., they should be cooled down.  Using a
proton (hydrogen atom) in the millikelvin range as an example, we
obtain an upper bound for $\epsilon$ to be $10^{-8}$, so
$\lambdabar_p/\epsilon\sim 100$\AA.  Thus, it is possible to have both
relatively sharp momenta and good localization, and so distinguishable
qubits can be created from elementary indistinguishable particles in
an appropriate momentum state. That is, a fiducial $N$-qubit product
state can be constructed from $N$ single-particle states as
\begin{equation}
  \label{eq:NQubitState}
  |\Psi \rangle_N = \otimes_{n=1}^N e^{-{\rm i}naP_z}|\Psi \rangle_1
   \, ,
\end{equation}
forming a one-dimensional lattice of particles with separation $a$. In
this case, we can loosely define a rest frame of these particles
(although they are not precisely in a zero momentum eigenstate), and
these particles are sufficiently distinguishable via their spatial
wavefunctions so that we can apply labels $1,\ldots,N$.  Thus, in
Alice's frame, the $N$ particles are prepared in a state where the
spatial wavefunctions of the particles are determined by the above
localization technique to make distinguishable qubits, but the spin
wavefunctions are completely arbitrary and can be used for encoding
quantum information.  In other inertial frames, these particles will
no longer be at rest but remain distinguishable.  From now on we
ignore the effects of momentum spread and consider the particles to be
eigenstates of momentum $\mathbf{p}$.

\section{Encoding in multiple particles}

Let Alice prepare $N$ particles in a state $|\Psi\rangle_N$ as
described above, choosing some arbitrary spin state, and consider the
state of these particles in Bob's reference frame. Let $T_N$ be the
(reducible) collective representation of the Lorentz group acting on
states of the $N$ particles, i.e., $T_N(\Lambda) =
T_1(\Lambda)^{\otimes N}$. A Lorentz transformation acts on the spin
state of each particle as a Wigner rotation via the SU(2)
representation $D^{1/2}$.  In fact, because these particles posses a
common momentum and they were all prepared with respect to a common
reference frame (Alice's), the group SU(2) acts \emph{identically} on
each spin via the reducible collective representation
$[D^{1/2}(\Omega)]^{\otimes N}$ for $\Omega \in$ SO(3).  If Bob does
not know the Lorentz transformation that relates his frame to Alice's,
then he represents the state of the $N$ particles as
\begin{equation}
  \label{eq:MixOverAllLorentzN}
  \mathcal{E}_N(|\Psi\rangle_N\langle\Psi|) = \int {\rm d}\Lambda\,
  f(\Lambda)T_N(\Lambda)|\Psi\rangle_N\langle\Psi| T_N(\Lambda)^\dag \, .
\end{equation}
We show that, for any prior distribution $f(\Lambda)$, there exists an
efficient encoding scheme that allows for quantum communication.  The
superoperator $\mathcal{E}_N$ has a decohering effect on the state of
the particles, but unlike~(\ref{eq:MixOverAllLorentz}) this
decoherence is not complete on the $N$-particle Hilbert space because
$T_N$ does not act irreducibly on the states of $N$ particles.
Because all the particles are now considered to have well-defined
momentum, the action on the reduced density operator $\rho_N$
describing the spin states of the $N$ particles is
\begin{equation}
  \label{eq:MixOverAllRot2}
  \mathcal{E}_N(\rho_N) = \int {\rm d}\Omega\,\tilde{f}(\Omega)
  [D^{1/2}(\Omega)]^{\otimes N} \rho_N [D^{1/2}(\Omega)^\dag]^{\otimes
  N} \, .
\end{equation}
where $\tilde{f}(\Omega)$ is induced by $f(\Lambda)$.  In the
following we assume the worst case scenario of a uniform prior
$\tilde{f}(\Omega)=1$.  Because $[D^{1/2}(\Omega)]^{\otimes N}$ acts
reducibly on the spin states, it is not completely decohering for
$N>1$.  By appealing to the techniques of decoherence-free
subspaces~\cite{Zan97} and noiseless subsystems~\cite{Kni00}, it is
possible use entangled states of multiple particles for encodings that
are completely protected against this form of decoherence.  Remarkably
(and conveniently), the noiseless subsystems for the superoperator
$\mathcal{E}_N$ are completely determined by the noiseless subsystems
for the spins under collective decoherence~\cite{Zan97,Kem01}, i.e.,
decoherence that acts identically on each particle.  The Hilbert space
of the $N$-particle spin states decomposes as
\begin{equation}
  \label{eq:DirectSum}
  \mathcal{H}_{j=1/2}^{\otimes N} = \bigoplus_{j=0}^{N/2}
  \mathcal{H}_{jR} \otimes \mathcal{H}_{jS} \, ,
\end{equation}
where SU(2) acts irreducibly on each subsystem $\mathcal{H}_{jR}$ (via
the irreducible representation of SU(2) labelled by $j$), and acts
trivially on the noiseless subsystems $\mathcal{H}_{jS}$.  Thus,
states encoded into a noiseless subsystem $\mathcal{H}_{jS}$ are
\emph{relativistically invariant}; they appear the same to all
inertial observers, regardless of their reference frame.  We note that
this encoding also protects against collective decoherence but is
still vulnerable to all other (standard) forms of decoherence, such as
the decay of the state $|\Psi \rangle_N$ via tunnelling of the
indistinguishable spin-$1/2$ particles.

The following example illustrates how a relativistically-invariant
qubit can be encoded into the state of four physical qubits.  Let four
particles be prepared in the spatial state as described above, making
them distinguishable, and let the spin states of these particles be
prepared in the $N=4$ singlet ($j=0$) subspace, which is
two-dimensional (i.e., an encoded qubit).
%\begin{align}
%  \label{eq:TwoSinglets}
%  |0_L\rangle &= \tfrac{1}{2}(|\!\uparrow\downarrow\rangle_{12} -
%   |\!\downarrow\uparrow\rangle_{12}) (|\!\uparrow\downarrow\rangle_{34} -
%   |\!\downarrow\uparrow\rangle_{34}) \\
%  \label{eq:Logical1}
%  |1_L\rangle &= \tfrac{1}{\sqrt{3}}(|\!\uparrow\uparrow\downarrow
%  \downarrow\rangle_{1234}  + |\!\downarrow\downarrow\uparrow
%   \uparrow\rangle_{1234}) \\
%  &\quad - \tfrac{1}{2\sqrt{3}}(|\!\uparrow\downarrow\rangle_{12} +
%   |\!\downarrow\uparrow\rangle_{12}) (|\!\uparrow\downarrow\rangle_{34} +
%   |\!\downarrow\uparrow\rangle_{34}) \, , \nonumber
%\end{align}
%where $\{|\!\!\uparrow\rangle,|\!\!\downarrow\rangle\}$ is \emph{any}
%orthogonal basis for the single qubit spin Hilbert space.
Because all states in this subspace possess zero total angular
momentum, the group of rotations acts trivially on this subspace.
Thus, the superoperator $\mathcal{E}_4$ preserves the two-dimensional
subspace spanned by these states, i.e., this subspace is
decoherence-free.  Encodings become more efficient for larger $N$, and
also if noiseless subsystems \cite{Kni00} (rather than subspaces) are
used.  Asymptotically, the number of logical qubits that can be
encoded into $N$ spin-1/2 particles in this manner is $N - \log_2
N$~\cite{Kem01}.

\section{Photons}

Much of the analysis for the massive particles applies to massless
photons as well, albeit with a different little group; thus, only the
key points of the photonic case will be mentioned.  The discrete
degrees of freedom for photons transform under a representation of the
little group for massless particles, and not under SU(2).  The
invariant subspaces under this group are the subspaces with zero
helicity.  Consider two entangled well-separated and therefore
distinguishable wave packets, with the same momentum profile centered
on $p$ (the construction for creating distinguishable qubits follows
the massive case). For example, the states
\begin{equation}
  \label{photon states}
  |\Psi_p^\pm \rangle =
  \tfrac{1}{\sqrt{2}}\bigl(|p,{+}\rangle|p,{-}\rangle \pm
  |p,{-}\rangle|p,{+}\rangle\bigr)\,,
\end{equation}
both satisfy $\mathbf{J}\cdot\mathbf{P} |\Psi_p^{\pm}\rangle=0$.  The
little group element for photons in the fiducial state $p^\mu =
(k,0,0,k)$ is decomposed~\cite{wei,nad1} as
\begin{equation}
  W(\Lambda,p)=S(\alpha,\beta)R_z(\omega(\Lambda,{\mathbf{p}})) \,,
\end{equation}
where $R_z(\omega)$ is a rotation by $\omega \in [0,2\pi)$ about the
$z$-axis and $S$ acts trivially on the physical states. The unitary
representation of the little group is just
\begin{equation}
  U_{\sigma
  \sigma'}(W(\Lambda,p))=e^{\text{i}\omega\sigma}\delta_{\sigma
  \sigma'}\,,
\end{equation}
where $\sigma=\pm 1$ denotes helicity. The states
transform as
\begin{equation}
  U(\Lambda)|p,\pm\rangle = \exp(\pm
  \text{i}\omega(\Lambda,{\mathbf{p}}))|\Lambda p,\pm\rangle\,,
\end{equation}
and so the encoded states $|\Psi_p^{\pm}\rangle$ will transform under
a general Lorentz transformation as
\begin{align}
  U(\Lambda)|\Psi_p^{\pm}\rangle &= \tfrac{1}{\sqrt{2}}
   \bigl(|\Lambda p,{+}\rangle_1|\Lambda p,{-}\rangle_2 \pm |\Lambda
   p,{-}\rangle_1|\Lambda p,{+}\rangle_2 \bigr) \nonumber \\
   &=|\Psi_{\Lambda p}^{\pm}\rangle \, .
\end{align}
%$U(\Lambda)|\Psi_p^{\pm}\rangle =|\Psi_{\Lambda p}^{\pm}\rangle$.
Thus one logical qubit can be encoded with two physical qubits
(photons) using the states $|\Psi_p^{\pm} \rangle$ as a basis.
Asymptotically, it is possible to encode $N - 2^{-1}\log_2 N$ qubits
in $N$ photons.  This encoding is analogous to the case of massive
particles with one direction shared between Alice and
Bob~\cite{Bar03}, which uses the noiseless subsystems that protect
against collective dephasing~\cite{Dua98}.

\section{Discussion}

The schemes presented for encoding quantum information into noiseless
subsystems are relativistically invariant because the encoded states
(in a noiseless subsystem $\mathcal{H}_{jS}$) are decoupled from any
degree of freedom associated with a reference frame (i.e., spatial and
angular momentum degrees of freedom).  States on the noiseless
subsystems $\mathcal{H}_{jS}$ describe entirely \emph{relative}
properties of the particles~\cite{Bar03b}, evidenced by the fact that
these subsystems carry irreducible representations of the symmetric
group for $N$ particles.

We note that the subsystems $\mathcal{H}_{jS}$ of the decomposition of
Eq.~(\ref{eq:DirectSum}) are only noiseless when the particles all
possess the same sharp momentum, because the Wigner rotation involved
is a function of the momentum.  However, these additional decoherence
effects (type-ii according to the classification of Sec.~I) are
typically small, of the order of $\Delta^2/\langle
E\rangle^2$~\cite{Per03}.

A key observation about this encoded relativistically invariant
quantum information is that it cannot be used directly for reference
frame distribution or alignment because of its fundamentally intrinsic
nature.  States suitable for reference frame alignment are not
invariant under reference frame transformations.  In current schemes
to perform such alignment, reference frames are encoded as
superpositions (over irreps $j$) of states on the subsystems
$\mathcal{H}_{jR}$ of Eq.~(\ref{eq:DirectSum})~\cite{Per01,Bag01} or
as superpositions of states entangled across the subsystems
$\mathcal{H}_{jR} \otimes \mathcal{H}_{jS}$~\cite{Chi04,Bag04}.
States encoded entirely in a noiseless subsystem $\mathcal{H}_{jS}$
(with the reduced state on $\mathcal{H}_{jR}$ completely mixed, say)
are invariant under reference frame transformations and therefore are
not suitable for alignment.  However, it is interesting to note that
Alice could prepare the system in a state of the form $\rho_{jR}
\otimes \sigma_{jS}$ on $\mathcal{H}_{jR} \otimes \mathcal{H}_{jS}$,
with directional information encoded in $\rho_{jR}$ (for the purposes
of reference frame alignment), and relativistically-invariant quantum
information encoded in $\sigma_{jS}$.  Bob can perform measurements of
linear and angular momentum on $\rho_{jR}$, obtaining information
about Alice's reference frame, without disturbing the encoded state
$\sigma_{jS}$.  For example, measuring the total linear momentum
provides information about the boost that relates Alice's frame to
Bob's, whereas performing measurements on the SU(2) representation
subsystems $\mathcal{H}_{jR}$ can provide information about the
orientation of Alice's frame relative to Bob's.  Thus, the
decomposition (\ref{eq:DirectSum}) of states of $N$ particles into
subsystems provides a division between states describing extrinsic
(spatial) and intrinsic properties.

Such encoded quantum information is, however, useful for most quantum
information processing tasks, such as quantum
teleportation~\cite{Ben93} of encoded states or demonstrating Bell's
theorem with observers who do not share a reference
frame~\cite{Bar03,Cab03}.  We also note that schemes for performing
quantum cryptography without a shared Cartesian frame (or in the
presence of noise)~\cite{Wal03,Boi03} can be extended in a
straightforward manner using the techniques here to perform quantum
cryptography between parties who do not share a Lorentz frame.

For quantum information processing, it is also necessary to
perform encoded logical operations.  Using the noiseless
subsystems for encoded states, the encoded operations are all
given by exchange interactions~\cite{Kem01}.  For elementary
spin-1/2 particles confined to a lattice as we describe, one would
naturally expect exchange interactions between the qubits; to
perform encoded operations, these interactions must be controlled
using electromagnetic fields. Finally, measurements may be
performed by performing projective measurements pairwise onto
singlet states.  For photons, recent progress in single photon
sources (c.f.~\cite{Vuc03}) may soon be able to create the
entangled encoded states of Eq.~(\ref{photon states}) with the
necessary wavepacket profiles and these advances give promise for
experimental realizations in the near future.

\begin{acknowledgments}
  We acknowledge significant contributions from Netanel Lindner, in
  particular on the photonic case, and thank Gerard Milburn, Terry
  Rudolph and Robert Spekkens for helpful discussions.  SDB would like
  to thank Enrique Solano and Frank Verstraete for highlighting the
  importance of establishing entanglement between observers in
  different Lorentz frames, and for valuable preliminary discussions
  on this subject.
\end{acknowledgments}

\end{document}